\documentclass[
reprint,
superscriptaddress,
amsmath,amssymb,
aps,
prl,
longbibliography
]{revtex4-2}

\usepackage[utf8]{inputenc}
\usepackage{amsfonts}
\usepackage{hyperref}
\usepackage{graphicx}
\usepackage{colortbl}
\usepackage{gensymb}

\usepackage[T1]{fontenc}
\usepackage{mathptmx}
\usepackage{bbold}
\usepackage{scalerel}
\usepackage{xcolor}
\usepackage{mathtools}

\usepackage{textcomp}

\begin{document}

\title{Shaping boundaries to control and transport topological defects in colloidal nematic liquid crystals}

\author{Gerardo Campos-Villalobos}
    \thanks{These authors contributed equally to this work.}
	\affiliation{Soft Condensed Matter \& Biophysics, Debye Institute for Nanomaterials Science, Utrecht University, Princetonplein 1, 3584 CC Utrecht, The Netherlands}
    \affiliation{CNR-ISC and Department of Physics, Sapienza University of Rome, p.le A. Moro 2, 00185 Roma, Italy}
\author{André F. V. Matias}
    \thanks{These authors contributed equally to this work.}
	\affiliation{Soft Condensed Matter \& Biophysics, Debye Institute for Nanomaterials Science, Utrecht University, Princetonplein 1, 3584 CC Utrecht, The Netherlands}
\author{Ethan I. L. Jull}
    \affiliation{Soft Condensed Matter \& Biophysics, Debye Institute for Nanomaterials Science, Utrecht University, Princetonplein 1, 3584 CC Utrecht, The Netherlands}
\author{Lisa Tran}
    \email{l.tran@uu.nl}
	\affiliation{Soft Condensed Matter \& Biophysics, Debye Institute for Nanomaterials Science, Utrecht University, Princetonplein 1, 3584 CC Utrecht, The Netherlands}
\author{Marjolein Dijkstra}
    \email{m.dijkstra@uu.nl}
    \affiliation{Soft Condensed Matter \& Biophysics, Debye Institute for Nanomaterials Science, Utrecht University, Princetonplein 1, 3584 CC Utrecht, The Netherlands}

\date{\today}

\begin{abstract}
Anisotropic rod-like particles form liquid crystalline phases with varying degrees of orientational and translational order. When confined geometrically, these phases can give rise to topological defects, which can be selected and controlled by tuning how the rods align near boundaries, known as anchoring. While anchoring in molecular liquid crystals can be controlled through surface functionalization, this approach is not easily applicable to microscale colloidal systems, which have so far been limited to planar anchoring. Here, using particle-based simulations, Landau-de Gennes theory, and experiments on colloidal rods, we demonstrate that topographical patterning of the boundary can effectively control the anchoring type and, in turn, the defect state in two-dimensional confined nematics. Building on this, we numerically predict that dynamically shape-shifting the boundaries can transform and transport topological defects.
\end{abstract}

\maketitle

In the 1940s, Onsager~\cite{onsager1949effects} predicted that sufficiently long rod-like particles exhibit a first-order phase transition from an isotropic phase, without positional and orientational order, to a nematic phase, with long-range orientational order but no positional order. In the bulk nematic phase, the long axes of the particles align along a common direction. When confined, as in twisted nematic display cells~\cite{castellano2005liquid}, boundaries disrupt the uniform nematic field, inducing elastic deformations of the director and increasing the free energy. Depending on the confinement geometry and the alignment of particles at the boundaries (anchoring), this frustration in the nematic field can lead to the spontaneous formation and stabilization of topological defects (TDs)~\cite{Mermin1979}. Formally, a TD corresponds to a configuration of the nematic field that cannot be continuously transformed into a uniform state. These defects may appear as points (hedgehogs), lines (disclinations), or walls (domain boundaries)~\cite{Alexander2012}. In 2D space, TDs reduce to point-like singularities where the orientation of the nematic director becomes ill-defined~\cite{Harth2020}. TDs are characterized by their topological charge $q$, which quantifies how the nematic director rotates around a closed loop encircling the defect~\cite{Kleman2003}. This charge determines the nature and stability of the defect, with different defect types corresponding to different values of $q$. Although TDs in 2D nematics are conventionally regarded as ``points'' interacting via elastic forces akin to the Coulomb force in electrostatics~\cite{Gennes1993}, recent evidence shows they also have intrinsic orientational properties that strongly influence their energetics and motion~\cite{vromans2016orientational,tang2017orientation}. One effective way to control the type and distribution of TDs is by manipulating the anchoring conditions of the liquid crystal under confinement~\cite{lavrentovich2003defects,shojaei2006role}. Adjusting the boundary conditions, such as \emph{planar} anchoring, where the director aligns tangent to the surface, or \emph{homeotropic} anchoring, where it aligns perpendicular, allows control over the symmetry, arrangement, and stability of TDs. This approach provides a versatile method for tailoring defect configurations, crucial in applications like photonic devices, where defect manipulation can control light propagation and enhance optical focusing~\cite{Hernandez2012}. In materials design, engineered defect patterns can be leveraged to fine-tune mechanical or optical properties~\cite{Lu2013,Khoo1993}. Additionally, in active matter systems, dynamic manipulation of defects can direct the collective behavior of self-organizing particles, enabling more predictable and controllable motion~\cite{Musevic2006}.

In recent years, significant progress has been made in controlling TDs via anchoring manipulation in molecular liquid crystals~\cite{Shin2018}. In these systems, altering the surface chemistry and composition allows for a fine control of anchoring conditions~\cite{Uline2010,Bryan-Brown1999,patel1993continuous,drawhorn1995anchoring}. However, much less has been achieved for larger-sized colloidal particles, where due to entropic arguments, planar anchoring is generally favored in the vicinity of flat and structured walls~\cite{van2000orientational,harnau2004colloidal,dijkstra2001wetting,lewis2014colloidal}. The anchoring direction can be controlled using grooved surfaces~\cite{Kim2016,Berreman1972}, however this preserves the planar anchoring and so achieving homeotropic anchoring of colloidal particles remains elusive, despite its fundamental importance~\cite{Babakhanova2019}. Recently, compelling evidence has been put forward that purely geometric control of boundaries in confined colloidal liquid crystals can steer the anchoring state and, consequently, the resulting defect structures~\cite{Kim2013}. Specifically, in 2D smectics of colloidal hard rods confined within elliptical cavities, the local radius of curvature determines the anchoring type. A transition from homeotropic to planar anchoring occurs at a \emph{critical} curvature that is approximately twice the length of the colloidal particles~\cite{jull2023curvature}. While these findings demonstrate how boundary curvature controls the alignment of anisotropic colloidal particles near curved regions, the influence of overall confinement anisotropy remains unclear, as does the applicability of these principles to geometries beyond simple elliptical shapes.

Inspired by surface functionalization in molecular liquid crystals, where anchoring is controlled by both energetic and entropic effects, we hypothesize that the anchoring of colloidal hard rods can be controlled purely through geometric boundary manipulations, driven solely by entropy. In this Letter, we use Monte Carlo (MC) simulations, Landau-de Gennes (LdG) theory, and experiments, to show that patterned boundaries alone can induce homeotropic anchoring in hard colloidal rods. We first demonstrate this entropically driven anchoring numerically and experimentally, then prove analytically that the anchoring transition arises from geometric constraints. Next, we show how different anchoring conditions impact the defect structure within circular confinements. Finally, we manipulate nematics under circular confinement with multiple anchoring conditions and apply this to control the charge and location of the topological defects.

\section{Results}

\begin{figure}[!t]
    \centering
    \includegraphics[width=.96\linewidth]{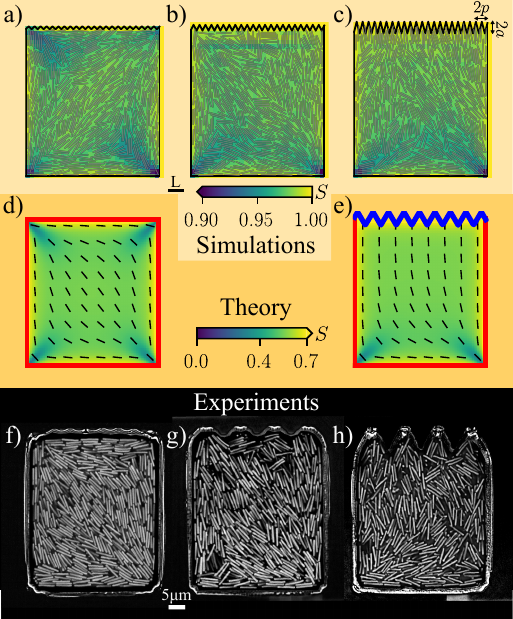}
    \caption[Nematic order of colloidal rods in rectangular confinement.]{\textbf{Nematic order of colloidal rods in rectangular confinement.} (a-c) Typical equilibrium configurations of rods confined to a rectangle with base $B/L = 9$ and height $A = B/0.9$, obtained from MC simulations, where the overlaid color map represents the scalar order parameter $S$. Scale bar is $L$. (d-e) Free-energy minimization using LdG theory, with color indicating the order parameter and black dashes showing the nematic director field. (f-h) Experimental results of silica rods. Scale bar is 5 {\textmu}m $\approx L$. The left column (a,d,f) corresponds to smooth boundaries ($a\approx p \approx0$) exhibiting planar anchoring, the middle column (b,g) shows results for an undulating top boundary ($a\neq$ $p\neq0$) with mixed anchoring, and the right column (c,e,h) with homeotropic anchoring.}
    \label{fig:FIGURE1}
\end{figure}

Using simulations, described in the Appendix, we first investigate how the boundary geometry affects the anchoring of colloidal rods, modeled as hard discorectangles with a length-to-diameter ratio of $L/D = 10$, confined within a rectangular container with base $B/L = 9$ and height $A = B/0.9$. We consider undulated boundaries, featuring a repeating series of sharp, acute-angled corners, which can locally direct the rod orientation at non-zero angles relative to the horizontal axis of the wall. We introduce these undulations along one of the short edges of the rectangle using a sinusoidal-triangular wave described by $y(x)=2a/\pi \arcsin\left[ \sin\left(2\pi x/p \right) \right]$, where $a$ and $p$ denote the amplitude and periodicity of the wave (in units of length), respectively, see Fig.~\ref{fig:FIGURE1}(c). The packing (area) fraction is fixed at $\eta=0.55$, where the 2D nematic phase is stable in bulk~\cite{bates2000phase}. We quantify local orientational order via a spatially resolved 2D nematic order parameter $S$~\cite{Grlea2016}. Consistent with previous work~\cite{lewis2014colloidal}, the equilibrium configuration of this system with nearly smooth boundaries ($a/D=1$ and $p/D=4$) features two disclination lines extending from opposite corners, each terminating in a TD with a charge of $q=-1/2$, as shown in Fig.~\ref{fig:FIGURE1}(a). This result is supported by a LdG theory for colloidal hard rods~\cite{everts2016landau,Gennes1993}, where free-energy minimization, as described in the Appendix, yields the equilibrium nematic director field shown in Fig.~\ref{fig:FIGURE1}(d). It also compares favorably with our experimental realization using silica rods~\cite{jull2023curvature,Kuijk2011,Kuijk2014} with a $L/D \simeq 9.4$ sedimented in 91 wt.-\% DMSO-in-water solution within a rectangular cavity fabricated on a \#1 cover glass (130-170 {\textmu}m thickness) using contact photolithography, as shown in Fig.~\ref{fig:FIGURE1}(f) and experimental Video~1, and described in the Appendix.

In MC simulations, going from a nearly flat wall, Fig.~\ref{fig:FIGURE1}(a), to a boundary with undulation parameters $a/D=4$ and $p/D=4$, Fig.~\ref{fig:FIGURE1}(c), we observe a clear transition from planar (tangential) anchoring to homeotropic (perpendicular) anchoring, which alleviates the frustration in rod orientation at the corners of the rectangular confinement, eliminating the disclination line along the undulated boundary. This behavior is confirmed by predictions from LdG theory~\cite{comsol}, Fig.~\ref{fig:FIGURE1}(e), and is also observed experimentally with silica rods, Fig.~\ref{fig:FIGURE1}(h) and experimental Video~2, where the persistence of homeotropic anchoring across imperfect undulating patterns and different $a/D$ and $p/D$ values suggests that the effect is robust across a range of parameters.

\begin{figure}[!t]
    \centering
    \includegraphics[width=.96\linewidth]{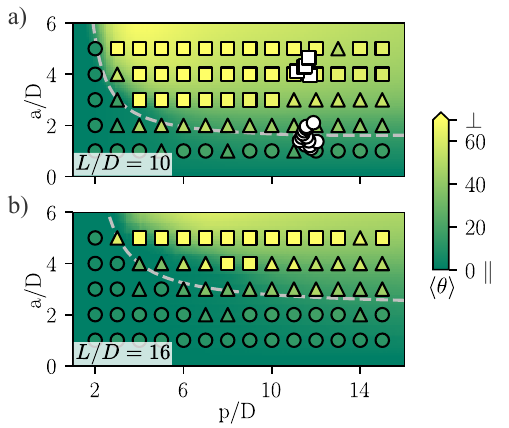}
    \caption[Rod anchoring.]{\textbf{Rod anchoring.} Color map of the average rod angle $\langle \theta \rangle$ relative to the horizontal axis as a function of the reduced undulation amplitude $a/D$ and period $p/D$. The background color represents $\langle \theta \rangle$ from numerical integration, and the colored symbols indicate $\langle \theta \rangle$ from MC simulations. Different symbols denote distinct anchoring types: planar (circles), homeotropic (squares) anchoring, and bi-stable states (triangles). White symbols indicate experimental measurements with corresponding anchoring types. The dashed gray line corresponds to Eq.~\eqref{eq:anchoring} with a $0.3$ prefactor. (a) Rod aspect ratio $L/D=10$ and (b) $L/D=16$.}
    \label{fig:FIGURE2}
\end{figure}

\begin{figure}[!h]
    \centering
    \includegraphics[width=0.98\linewidth]{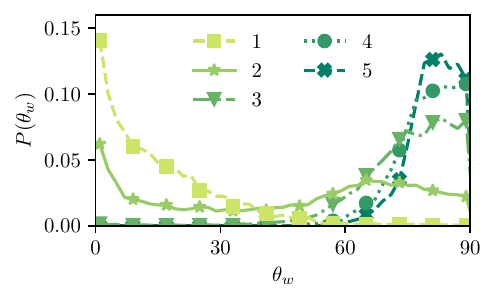}
    \caption[Rod alignment probability distribution.]{\textbf{Rod alignment probability distribution.} Probability distribution function $P(\theta)$ of the rod angle for various undulation amplitudes at fixed $p/D=4$. For $a/D=1$, most rods align near $0\degree$ (planar anchoring), for $a/D>2$, they align near $90\degree$ (homeotropic anchoring), and at $a/D=2$, the rod orientation is bi-modal with peaks at $0\degree$ and $90\degree$, indicating a bi-stable state.}
    \label{fig:histogram}
\end{figure}

\subsection{Anchoring as a function of pattern parameters}

To test this hypothesis, we systematically vary $a$ and $p$ in simulations, fixing $B/L = 10$ and $A = 2B$, and additionally consider longer rods with $L/D = 16$. In MC simulations, we determine the anchoring type by measuring the average angle $\langle \theta \rangle$ between the long axis of the rods near the patterned boundary and the horizontal axis. In addition, we also numerically integrate the average angle of a single rod near an undulating boundary. We first check whether the rod is within the boundary when its center is positioned at $(x,y)$ with an orientation $\theta$. We vary $x \in [0,p)$, $y \in (-2a-D,0]$, and $\theta\in[0,\pi)$, where $a$ and $p$ denote the amplitude and periodicity of the undulation, and $(x,y)=(0,0)$ corresponds to the top-left corner of the undulating pattern. The average angle is then computed by averaging over all accepted positions and orientations. The average angle predicted by the single-rod calculations agrees well with the results from MC simulations, as seen in Fig.~\ref{fig:FIGURE2}. However, in the homeotropic region ($a\gtrsim 3$ and $p\gtrsim 2$), the average angle is smaller than in the MC simulations, resulting in a transition at a smaller value $\langle \theta \rangle \approx 35\degree$. This discrepancy arises because, in the MC simulations, rod-rod interactions in the oriented nematic phase enhance the tendency for perpendicular alignment to the wall.

For low values $\langle \theta \rangle \approx 0 \degree$, where the boundary is too shallow or the peaks and valleys are too close, most rods align tangent to the boundary (planar anchoring), while $\langle \theta \rangle \approx 90\degree$ indicates that most rods are perpendicular to the boundary (homeotropic anchoring). In Fig.~\ref{fig:FIGURE2}, the colored markers represent the average rod angle $\langle \theta \rangle$ from MC simulations. Different markers indicate distinct anchoring types: circles correspond to planar anchoring ($\langle \theta \rangle < 45\degree$), while squares represent homeotropic anchoring ($\langle \theta \rangle > 45\degree$). This trend is also confirmed experimentally, as shown by the white symbols in Fig.~\ref{fig:FIGURE2}(a). Because of the resolution limits of soft lithography, the fabricated boundaries exhibit smoothed features and only a limited range of $a/D$ and $p/D$ values have been explored experimentally.

\begin{figure*}[!t]
	\centering
	\includegraphics[width=\linewidth]{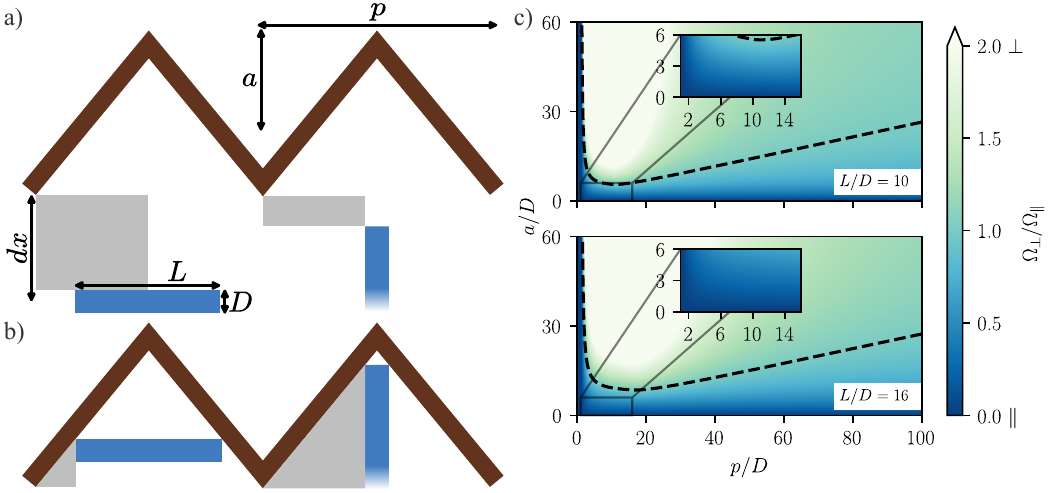}
	\caption[Rod alignment accessible area.]{\textbf{Rod alignment accessible area.} (a-b) Schematic of the accessible areas (gray) for a rod with tangential (left column) and perpendicular (right column) orientation. When the rod is located inside the undulating pattern (b), only the gray triangular region is accessible. (c) Color map of the ratio between the areas accessible to perpendicular and tangential alignment $\Omega_\perp/\Omega_\parallel$ as a function of the reduced undulation amplitude $a/D$ and period $p/D$ for different $L/D=10$. Blue indicates regions where the accessible area for tangential alignment is larger ($\Omega_\parallel > \Omega_\perp$), while light green corresponds to regions where perpendicular alignment is favored. The insets correspond to the parameter region explored in Fig.~\ref{fig:FIGURE2}. The dashed lines correspond to the transition from tangential to perpendicular alignment ($\Omega_\parallel / \Omega_\perp = 1$).}
	\label{fig:anchoring_strenght_end}
\end{figure*}

In between we observe bistability, represented by triangles, as a result of two effects: (i) instantaneous spatial heterogeneities, where regions with small and large rod angles coexist, as seen in Fig.~\ref{fig:FIGURE1}(b,g), simulation Video~3 and experimental Video~4, and (ii) dynamic large rod domains that collectively switch between the two anchoring states, as shown in simulation Video~3. The anchoring bistability can be characterized by the probability density function $P(\theta)$ shown in Fig.~\ref{fig:histogram}. For $a/D=1$ (light green squares), most rods are tangent to the boundary, indicating planar anchoring. As the amplitude $a/D$ increases (for example, dark green crosses), the distribution shifts toward larger angles, indicating a transition toward homeotropic anchoring. Near the transition, at $a/D=2$ (green stars), the distribution becomes bimodal, providing evidence of bistability in the rod orientation. This behavior is expected, as Landau-de Gennes calculations indicate that the relative free-energy difference between the two states is only about $5\%$.

\subsection{Entropy driven anchoring}

Since the rods interact purely via excluded-volume interactions, both with each other and with the boundary, the anchoring behavior can be rationalized solely through entropic considerations. To explore this, we neglect interactions between rods and determine analytically the probability of a rod adopting homeotropic or planar alignment based on the geometrical framework presented in Ref.~\cite{king2023bobas}. In that work, a toy model of a single colloidal rod was used to determine the packing fractions at which the transition from isotropic to nematic and smectic phases occurs. Here, we adapt their toy model to explore how the undulation parameters, $a$ and $p$, influence the emergence of homeotropic anchoring. To this end, we calculate the free energy per test particle using Gibbs's definition
\begin{equation}
	\beta F = \int_V d\mathbf{r} \sum_\mathcal{T} P\left(\mathbf{r}, \mathcal{T}\right) \log P\left(\mathbf{r}, \mathcal{T}\right) \\,
\end{equation}
where $\beta=(k_BT)^{-1}$ is the inverse temperature, $P\left(\mathbf{r}, \mathcal{T}\right)$ is the probability of the test particle exhibiting anchoring type $\mathcal{T}$ at position $\mathbf{r}$. The sum runs over all possible anchoring types, and the integral is taken over the entire accessible volume $V$. We simplify the model by assuming that rods with homeotropic anchoring lie perfectly perpendicular to the boundary, with probability $p_\perp$, while rods with planar anchoring lie perfectly tangent to the boundary, with probability $p_\parallel=1-p_\perp$. Under this assumption, the probability of finding a perpendicularly aligned rod at position $\mathbf{r}$ is given by
\begin{equation}
	P\left(\mathbf{r}, \perp\right) = \frac{p_\perp}{\Omega_\perp}\omega_\perp(\mathbf{r}) \\,
\end{equation}
and similarly, the probability of finding a tangentially aligned rod at position $\mathbf{r}$ is
\begin{equation}
	P\left(\mathbf{r}, \parallel\right) = \frac{1 - p_\perp}{\Omega_\parallel}\omega_\parallel(\mathbf{r}) \\,
\end{equation}
where $\omega_\mathcal{T}$ denotes the spatial weights that pick out the allowed positions $\mathbf{r}$ of a test rod with type $\mathcal{T}$ orientation, and $\Omega_\mathcal{T}$ represents the normalization constant
\begin{equation}
	\Omega_\mathcal{T} = \int_V d\mathbf{r} \omega_\mathcal{T}(\mathbf{r}) \\,
\end{equation}
corresponding to the accessible area for type $\mathcal{T}$ alignment. To simplify the expression for the free energy, we take advantage of the fact that $\int_V \omega_\mathcal{T}\log\left(\omega_\mathcal{T}\right)=0$, which allows us to combine the previous expressions into
\begin{equation}
	\beta F = \beta F_0(p_\perp) - p_\perp\log \Omega_\perp - (1 - p_\perp)\log \Omega_\parallel \\,
	\label{eq:freeEnergy}
\end{equation}
where $\beta F_0(p_\perp) = p_\perp\log p_\perp + (1 - p_\perp)\log\left(1 - p_\perp\right)$ corresponds to the standard entropy of mixing for the two anchoring types.

To determine the most likely rod alignment, we minimize the free energy with respect to the alignment probability $p_\perp$
\begin{equation}
	\frac{\partial \beta F}{\partial p_\perp} = \log\left(\frac{p_\perp}{1 - p_\perp}\right) - \Delta S \\,
\end{equation}
where $\Delta S = \log(\Omega_\perp / \Omega_\parallel)$ quantifies the entropic preference for one alignment over the other. Setting the derivative to zero yields the equilibrium probability.
\begin{equation}
	p^*_\perp = \frac{1}{1 + e^{-\Delta S}} \\.
\end{equation}
When both accessible areas are equal, \textit{i.e.} $\Delta S=0$, the probabilities of a rod aligning perpendicular or tangential to the wall are identical. Deviations from this condition indicate a preference for one type of anchoring: a positive $\Delta S$ favors homeotropic alignment, while a negative $\Delta S$ favors planar alignment. Consequently, the ratio $\Omega_\perp / \Omega_\parallel$ serves as an indicator of the effective strength of homeotropic anchoring $w$, while its inverse reflects the strength of planar anchoring.

\begin{figure*}[!t]
	\centering
	\includegraphics[width=\linewidth]{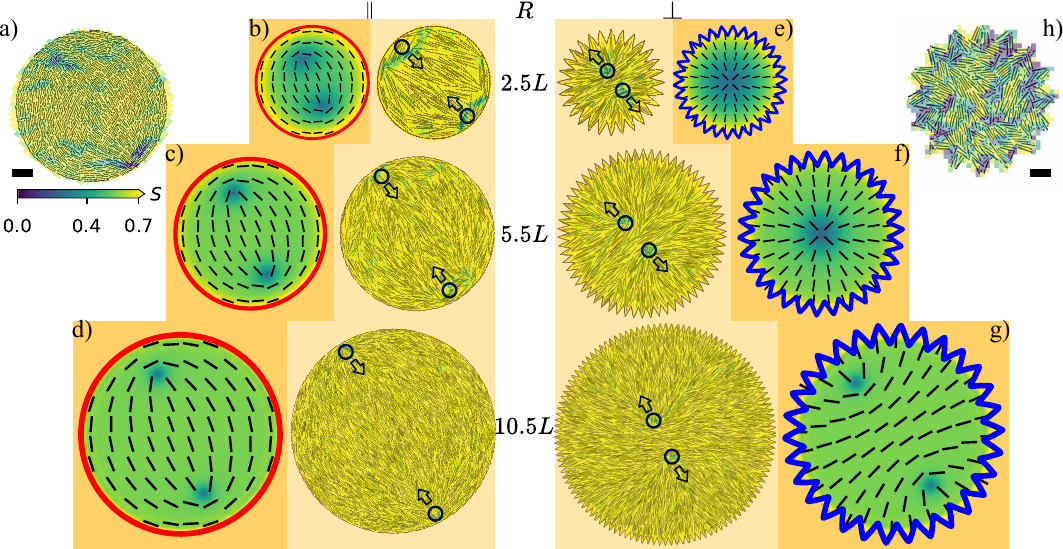}
	\caption[Topological defects in circular confinement]{\textbf{Topological defects in circular confinement.} Topological defects in a nematic state confined within a circle of radii $R=2.5L$, $5.5L$, and $10.5L$, with smooth (a-d) and undulated boundaries (e-h). The color indicates the local nematic order. Black circles mark the approximate positions of the TDs. Colored background frames distinguish between LdG theory (dark yellow), and MC simulations (light yellow).}
    \label{fig:FIGURE3}
\end{figure*}

To estimate the anchoring strength, we determine the accessible areas $\Omega_\perp$ and $\Omega_\parallel$ for perpendicular and tangential alignment, respectively. The total accessible area is decomposed into two regions: a rectangular region corresponding to configurations where the rod lies outside the undulating pattern, and a triangular region where the rod is located within it, see Fig.~\ref{fig:anchoring_strenght_end}(a-b). We begin by considering a rod with tangential alignment; in this case, the accessible area is given by
\begin{equation}
	\Omega_\parallel = \frac{p}{2} \left(dx - \frac{D}{2}\right) + \frac{1}{2}\left(p - L\right)^2 \frac{a}{p} \Theta\left(1 - \frac{L}{p}\right) \\,
\end{equation}
where the first term corresponds to the accessible area outside the undulating pattern, Fig.~\ref{fig:anchoring_strenght_end}(a), and the second term accounts for the area inside the undulation, Fig.~\ref{fig:anchoring_strenght_end}(b), which only contributes if the undulation period $p$ exceeds the rod length $L$. This condition is enforced using the Heaviside function $\Theta$, which equals zero when its argument is negative and one otherwise. Similarly, for a rod with perpendicular alignment, the accessible area is given by
\begin{equation}
	\Omega_\perp = \frac{p}{2} \left(dx - \frac{L}{2}\right) + \frac{1}{2}\left(p - D\right)^2 \frac{a}{p} \Theta\left(1 - \frac{D}{p}\right) \\,
\end{equation}
where the second term becomes non-zero when $p>D$. In Fig.~\ref{fig:anchoring_strenght_end}(c), we plot the accessible area ratio $\Omega_\perp/\Omega_\parallel$ for rod aspect ratio $L/D=10$. A ratio larger than one (shown in light green) indicates a preference for perpendicular alignment, \textit{i.e.} homeotropic anchoring, while a ratio less than one (blue) indicates a preference for tangential alignment, \textit{i.e.} planar anchoring. Furthermore, we determine analytically the boundary between these two anchoring regimes, shown as dashed line in Fig.~\ref{fig:anchoring_strenght_end}(c). For example, in the case $ D < p < L $, the anchoring transition boundary is given by
\begin{equation}
	\frac{\Omega_\perp}{\Omega_\parallel} = \frac{2a}{L-D}\left(1 - \frac{D}{p}\right)^2 = 1 \\.
    \label{eq:dS_intermidiate}
\end{equation}
Note that this result is independent of $dx$.

The transition between between planar and homoetropic anchoring, shown in Fig.~\ref{fig:anchoring_strenght_end}(c), deviates from the Monte Carlo prediction. However, like on the previous case of the angle of single rod, the discrepancy arises because we neglect rod-rod interactions, and so, by biasing the accessible area ratio towards the homeotropic anchoring, $\Omega_\perp / \Omega_\parallel = 0.3$, we get a boundary line that corresponds closely to the anchoring transition observed in the numerical results, as seen by the gray line in Fig.~\ref{fig:FIGURE2}. Thus, we hypothesize that the anchoring strength $w$ is proportional to
\begin{equation}
	\frac{w}{L} \propto \frac{2a}{L-D}\left(1 - \frac{D}{p}\right)^2 \\.
    \label{eq:anchoring}
\end{equation}

\begin{figure*}
    \centering
    \includegraphics[width=\linewidth]
    {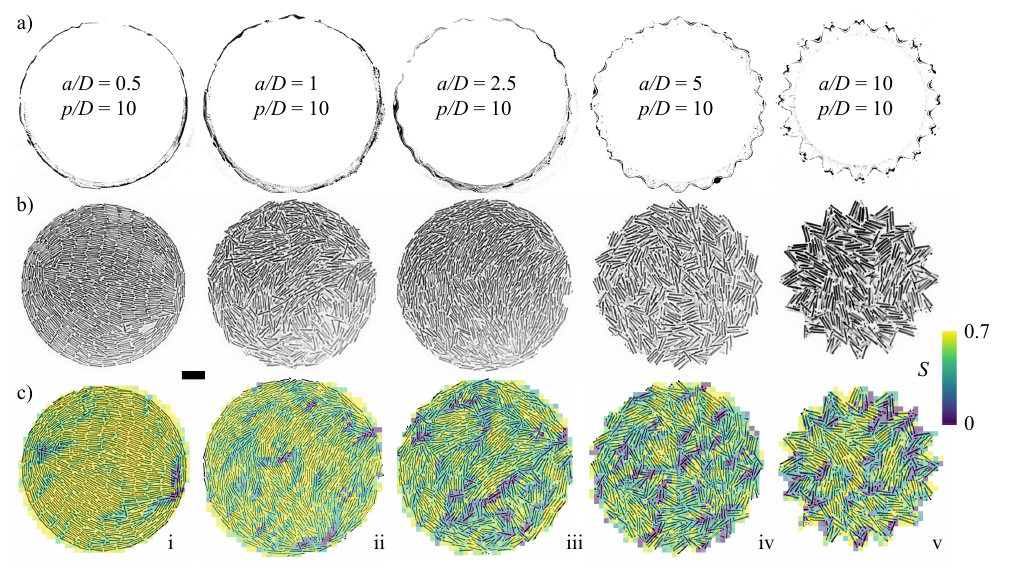}
    \caption[Defect structures for varying boundary topography in experiments]{\textbf{Defect structures for varying boundary topography in experiments.} Scale bar is 10 {\textmu}m. (a) Outlines of confinement wells fabricated via contact photolithography. The confinement walls are imaged in reflection, and the intensity is thresholded to highlight the wall geometry. The amplitude $a$ and period $p$ (normalized by the rod diameter $D$) correspond to the values on the photomask. Due to the finite lithographic resolution, the actual topographical profiles are slightly smoothed, resulting in a reduced $a/D$ in the fabricated wells. (b) Confocal micrographs of fluorescent silica rods sedimented into the confinement wells of (a) with varying topographies. (c) Defect configurations identified from the confocal micrographs in (b) using a color map of the local nematic order parameter $S$. With increasing $a/D$ from i-v, the confinement walls promote stronger homeotropic anchoring, leading to a higher number of defects in the center of the system.}
    \label{fig:experimentfigure1}
\end{figure*}

This agreement between MC simulations, numerical integration, and the analytical calculation also holds for the case of longer rods with a larger aspect ratio, $L/D=16$, see Fig.~\ref{fig:FIGURE2}(b).

\subsection{Circular confinement}

We now turn to circular confinement to explore how anchoring alone can be leveraged to manipulate TDs. We consider circular cavities with radii between $R=2.5L$ and $10.5L$. To facilitate TD identification we consider longer rods, $L/D=25$. We examine two boundary conditions, planar and homeotropic anchoring, $a/D=10$ and $p/D=14$. In Fig.~\ref{fig:FIGURE3}(b-g), we present the typical defect structures for different radii obtained from our LdG theory and observed in our MC simulations. Consistent with previous studies~\cite{delasHeras2009}, nematics in circular confinement with planar anchoring adopt a bipolar configuration with two diametrically-opposed TDs, each with a charge $q=+1/2$, Fig.~\ref{fig:FIGURE3}(b-d). For homeotropic anchoring, Fig.~\ref{fig:FIGURE3}(e-g), the MC simulations reveal a similar antipodal defect configuration. However, in this case, the orientation of the comet-like fractional TDs is reversed, and the average defect separation is consistently smaller than in the planar anchoring case. Although the bipolar configuration remains predominantly preserved, for $R=5.5L$, the increased geometric frustration leads to transient fluctuations, where the two $q=+1/2$ defects cyclically merge into a single defect with $q=+1$, pinned near the circle center. This behavior is shown in simulation Video~5 for rods with $L/D=10$. This is due to the comparable free energy of these configurations (with a relative difference of $\approx 5\%$), as confirmed by LdG theory. As shown in Fig.~\ref{fig:FIGURE3}, the single-defect becomes energetically unfavorable at larger radii, leaving the two-defect state as the only one observed configuration.

Experimental results in Fig.~\ref{fig:FIGURE3}(a,h) and Fig.~\ref{fig:experimentfigure1} further demonstrate how topographical anchoring control can alter defect states, from boojums under planar anchoring to bulk disclinations with homeotropic anchoring. The experimental circular confinement wells shown in Fig.~\ref{fig:experimentfigure1}(a) were produced via contact photolithography and have topographies with a fixed periodicity of $p/D=10$ and five different amplitudes of $a/D=0.5$, 1, 2.5, 5, and 10. The rod packing fraction $\eta$ is low enough to access both the nematic and smectic phase for our rods with aspect ratio $L/D\approx 9$~\cite{bates2000phase}. The average value of $S$ in the different confinement wells is represented by a color map, shown in Fig.~\ref{fig:experimentfigure1}(b). As in the simulations, increasing $a/D$ from 0.5 in Fig.~\ref{fig:experimentfigure1}(b)-i to 10 in Fig.~\ref{fig:experimentfigure1}(b)-v, shifts the anchoring from planar to homeotropic. With planar anchoring in Fig.~\ref{fig:experimentfigure1}(b)-i, two boojum defects are formed at or near the boundary, marked by low $S$ values in the color map, characteristic of nematics in circular confinements with planar anchoring. As the homeotropic anchoring strength increases, Fig.~\ref{fig:experimentfigure1}(b)-ii to Fig.~\ref{fig:experimentfigure1}(b)-v, more and more defects nucleate within the bulk and far from the walls, yielding defect states more complex than the purely nematic systems studied by LdG theory and MC simulations. Given the presence of smectic ordering and the long relaxation times of large colloidal assemblies, the large number of disclinations for homeotropic systems is unsurprising. Still, these results demonstrate the effectiveness of boundary topography in altering colloidal liquid crystal anchoring to tune the defect state.

\begin{figure}[!t]
    \centering
    \includegraphics[width=.96\linewidth]{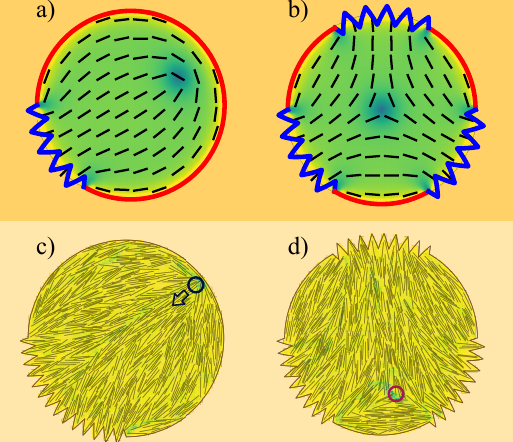}
    \caption[Defect control of nematics under circular confinement.]{\textbf{Defect control of nematics under circular confinement.} (a,c) Nematics confined within a circle with planar anchoring and a single $60\degree$ slice of homeotropic anchoring results in a $+1/2$ defect (black circle and arrow) positioned opposite to the homeotropic region. (b,d) Nematics confined within a circle with alternating homeotropic and planar anchoring in $60\degree$ slices produces a central $-1/2$ defect (red circle). Colored background frames distinguish between LdG theory (dark yellow), and MC simulations (light yellow).}
    \label{fig:defect_circle}
\end{figure}

\begin{figure*}[!t]
	\centering
	\includegraphics[width=\linewidth]{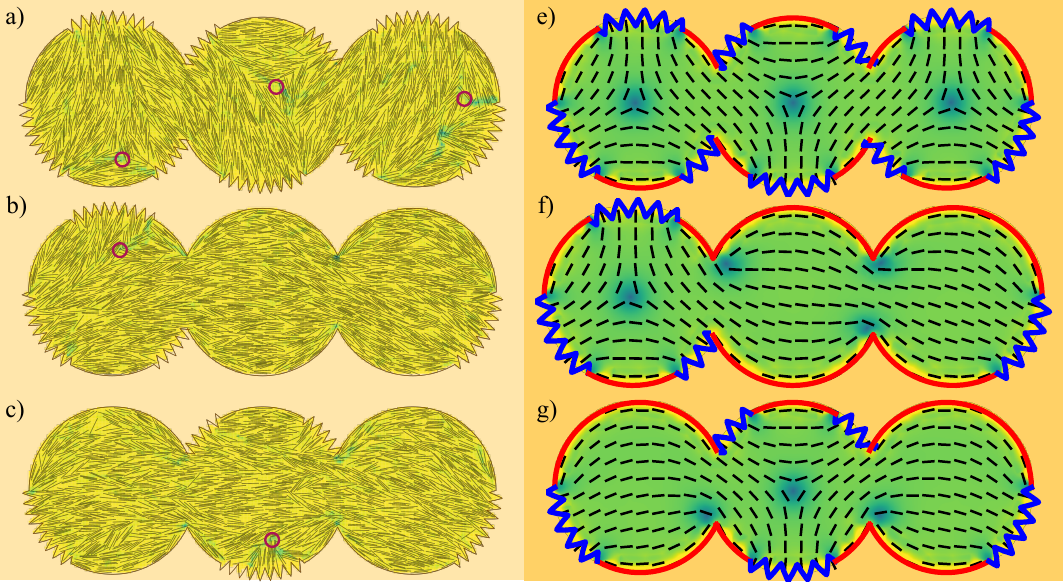}
	\caption[Defect-based information storage.]{\textbf{Defect-based information storage.} Nematics confined within three slightly overlapping circular confinements with different anchoring patterns, where a $-1/2$ defect is labeled as \texttt{1} and no defect as \texttt{0}. The defect can be placed in all three circles, \texttt{111} (a,e), in an edge circle, \texttt{100} (b,f), or in the central circle, \texttt{010} (c,g). Colored background frames distinguish between LdG theory (dark yellow), and MC simulations (light yellow).}
    \label{fig:FIGURE4}
\end{figure*}

\begin{figure}[!t]
	\centering
	\includegraphics[width=\linewidth]{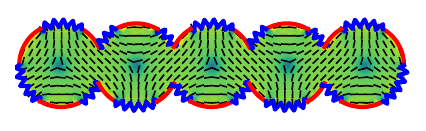}
	\caption[Defect manipulation in five circles]{\textbf{Defect manipulation in five circles.} Nematics confined within five slightly overlapping circular confinements with different anchoring patterns store 5 bits of information as obtained from Landau-de Gennes free-energy minimization. Here, a $-1/2$ defect is placed in all five circles, corresponding to a \texttt{11111} configuration.}
	\label{fig:5_circles}
\end{figure}

Finally, we explore the active control of nematic defect structures, demonstrating that they can be reversibly reconfigured by modifying the boundary conditions. We focus on circular confinement where anchoring can be selectively patterned in $60\degree$ slices, Fig.~\ref{fig:defect_circle}. When a single $60\degree$ slice imposes homeotropic anchoring, a single $+1/2$ defect emerges opposite to the homeotropic boundary as shown in Fig.~\ref{fig:defect_circle}(a,c). The remaining TD is now distributed along the boundary discontinuities, with each contributing a winding number of $+1/4$, ensuring TD conservation~\cite{Yao2022}. In contrast, when the $60\degree$ slices alternate between planar and homeotropic anchoring, shown in Fig.~\ref{fig:defect_circle}(b,d), six boundary discontinuities are present, totaling a winding number of $+3/2$. Thus, charge conservation results in a $-1/2$ defect near the center of the circle. With this strategy we can precisely control defect configurations in nematics confined within multiple cavities where a $-1/2$ defect is labeled as \texttt{1}, and its absence as \texttt{0}. To illustrate this, we consider three slightly overlapping circular cavities of radius $R=5L$, with centers spaced $1.8R$ apart. When all circles have alternating boundary conditions, each circle hosts a $-1/2$ defect, \texttt{111} state, Fig.~\ref{fig:FIGURE4}(a,e). By modifying the boundary conditions, the central TDs vanish, as seen in Fig.~\ref{fig:FIGURE4}(b,f) and (c,g), allowing independent control over all three TDs. These TDs are not only predicted by LdG theory but also reproduced in MC simulations, with thermal fluctuations evident in the patterns shown in Fig.~\ref{fig:FIGURE4}(a-c). We emphasize that, as before, the circle radius must be small enough to allow for single-defect configurations. Furthermore the spacing between circle centers is crucial: at smaller distances, TDs are not fully isolated, while at larger distances, the system behaves as multiple independent units. While we illustrate this concept using three circles, the approach can be readily extended to an arbitrary number of circles (Fig.~\ref{fig:5_circles}). Beyond controlling the TD position, an intrinsic free-energy path connects different defect locations. Thus, in a dynamic system with shape-shifting boundaries, defects can be actively transported between adjacent circle centers, as shown in simulation Video~6. To have a clear free-energy path, we increased the chemical potential ($\beta \mu = 5$), elastic constant ($l_s=50\times 0.837L^2$), and anchoring strength ($w=10^3L$) in the LdG theory. Given that colloidal particles can be ``trapped'' by topological defects~\cite{Martinez2011}, shape-shifting boundaries could also offer a powerful mechanism for controlled transport of colloids.

In conclusion, we have demonstrated, by MC simulations, LdG theory, and experiments, that the anchoring type, and consequently the defect state, of confined colloidal hard rods can be precisely controlled by employing topographically patterned, undulated boundaries. We have mapped out the geometric parameters of patterned boundaries that govern the entropy-driven transition between planar and homeotropic anchoring. Interestingly, we identified an intermediate bistable region in which rod-like particles exhibit no clear preference for either anchoring type. Our analysis, based on intuitive geometric and entropic considerations, reveals that anchoring strength depends not only on these boundary parameters but also on particle dimensions such as rod length and diameter. Although not yet realized experimentally, our MC simulations and LdG theory predict that dynamically shape-shifting boundaries facilitate controlled transformations and transportation of TDs, allowing precise control over their structure and position. We find that the creation and motion of TDs is facilitated under conditions of low chemical potential and/or high elastic constants. We anticipate that this study can be extended to thermotropic liquid crystals, where geometrically patterned surfaces with spatially varying anchoring conditions could be realized through localized surface functionalization. Furthermore, we hope this work motivates experimental realizations of shape-shifting boundaries, thereby opening new avenues for defect transport and information storage~\cite{kos2022nematic}. This could potentially be achieved through the use of responsive materials such as liquid crystal elastomers~\cite{Warner2003}. In particular, chiral liquid crystal elastomers are especially promising for creating dynamic, actuable topographies with controllable periodicity (\textit{e.g.} by patterning topological defects in the elastomer~\cite{Feng2018} and by varying the elastomer pitch~\cite{Peixoto2022}).

\section*{Acknowledgments}
G. C.-V. adn A.F.V.M. thank Rodolfo Subert for useful discussions and acknowledges funding from The Netherlands Organization for Scientific Research (NWO) through the ENW PPS Fund 2018 - Technology Area Soft Advanced Materials (ENPPS.TA.018.002). A.F.V.M and M.D. acknowledge funding from the European Research Council (ERC) under the European Union’s Horizon 2020 research and innovation program (Grant Agreement No. ERC-2019-ADG 884902 SoftML) and partial support from the European Union’s Horizon Europe research and innovation program under the grant agreement number 101203506, Marie Sklodowska-Curie Action Postdoctoral Fellowship, project IonFlowElast. E.I.L.J. acknowledges funding from the European Commission (Horizon-MSCA, Grant No. 101065631). L.T. acknowledges support from the European Commission (Horizon-MSCA, Grant No. 892354) and the NWO ENW Veni grant (Project No. VI.Veni.212.028). E.I.L.J. and L.T. acknowledge support from the Starting PI Fund for Electron Microscopy Access from Utrecht University’s Electron Microscopy Center.

\section*{Data Availability}
The data that support the findings of this article, the code used to analyze it, and the source code and parameters used to generate the simulations are publicly available~\cite{campos_villalobos_2026_19846577}.

\section{Appendix}

\subsection{Experiments}

\begin{figure}[!t]
	\centering
	\includegraphics[width=0.85\linewidth]
    {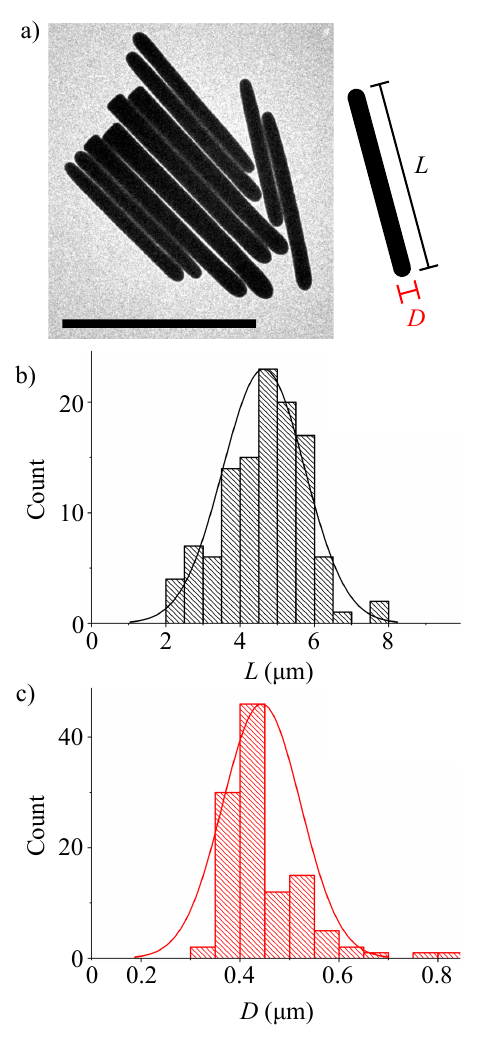}
    \caption[Characterization of synthesized silica rods]{Characterization of synthesized silica rods. (a) Representative transmission electron micrograph of synthesized silica rods. The scale bar is 5 {\textmu}m. Polydispersity in rod length $L$ and diameter $D$ is measured across several micrographs. (b,c) Plots of $L$ and $D$ of 115 silica rods. The average rod length is $L=4.6 \pm 0.1$ {\textmu}m, with a polydispersity index of 24\%. The average rod diameter is $D=443 \pm 7$ nm, with a polydispersity index of 18\%.}
    \label{fig:experiment2}
\end{figure}

In our experiments, following protocols from our prior work \cite{jull2023curvature}, we synthesized fluorescently-labeled silica rods that are then characterized using transmission electron microscopy, Fig.~\ref{fig:experiment2}(a), and imaged in confinement using high-resolution, confocal laser scanning microscopy, Fig.~{1}(g-h). The silica rods were prepared using the method by Kuijk \textit{et al.}~\cite{Kuijk2011}. $30.1$ g polyvinylpyrrolidone (PVP, MW 40,000, Sigma-Aldrich) was fully dissolved in $300$ mL 1-pentanol (${\geq}99{\%}$, Honeywell) with sonication and stirring. To this, $30$ mL ethanol (Baker), $10$ mL ultrapure water (Millipore), and $2$ mL $0.19$ M sodium citrate dihydrate (${\geq}99{\%}$, Sigma-Aldrich) were added. After adding $6.75$ mL ammonium hydroxide (${\geq}25{\%}$, Sigma-Aldrich), the mixture was shaken to form a micro-emulsion, then left for $120$ seconds to stabilize. $12$ mL TEOS (Sigma-Aldrich) was introduced in four $3$ mL doses every 6 hours. The mixture was gently inverted after each addition. After the final dose, the solution reacted for 24 hours.

Fluorescent labeling was done per Kuijk \textit{et al.}~\cite{Kuijk2014}. $9.5$ mg RITC dye (Sigma) was dissolved in $1.34$ mL ethanol, followed by $9.5$ {\textmu}L APTES (${\geq}98{\%}$, Sigma-Aldrich) and left overnight to react. About $0.02$ g of silica rods were dispersed in $100$ mL ethanol with $5$ mL each of ammonium hydroxide and water. Three additions of $10.3$ {\textmu}L TEOS and $53.3$ {\textmu}L RITC-APTES were made at 60-minute intervals (totaling $31$ and $160$ {\textmu}L, respectively). To grow an outer silica layer, labeled rods were placed in $300$ mL ethanol, $10$ mL water, and $10$ mL ammonium hydroxide, and three $9.52$ {\textmu}L TEOS additions were made hourly. Each shell growth step was left overnight to react.

Our fluorescently-labeled silica rods have an average length of $L{=}4.6{\pm}0.1$ {\textmu}m with a polydispersity of $\sim 24\%$ and a diameter of $D{=}443{\pm}7$ nm with a polydispersity of $\sim 18\%$, Fig.~\ref{fig:experiment2}(b), and Fig.~\ref{fig:experiment2}(c)). Averaging over the length-to-diameter aspect ratios of each measured rod gives $L/D{\approx}9.4{\pm}0.1$, with an aspect ratio polydispersity of $\sim 15\%$.

To match the refractive index of silica microparticles ($\sim1.45$), the rods are sedimented in 91 wt.-\% dimethylsulfoxide (DMSO, Sigma-Aldrich) in DI water. At the bottom of our samples are confining wells that are produced on ${\#}1$ cover glass ($130$-$170$ {\textmu}m thickness, VWR) using contact photolithography, following the protocol of our prior work~\cite{jull2023curvature}. In brief, SU-8 2010 (Micro Resist Technology) was spin-coated at $300$ rpm ($10$ s) and $1500$ rpm ($30$ s) to form a $\sim10$ {\textmu}m layer. Edge bead removal was performed, followed by a soft bake at $95~^{\circ}$C for 3 min. A chrome photomask (JD PhotoData) was aligned for UV exposure ($210$ mJ/cm$^2$ at $30$ mW/cm$^2$, $30\%$) using a UV-EXP150R system (idonus s\`{a}rl). After a 4 min post-exposure bake at $95~^{\circ}$C, samples were developed in mr-Dev 600 (Micro Resist Technology) for 10 min, rinsed with isopropyl alcohol, and hard-baked at $200~^{\circ}$C for 1 hour. Notably, the produced topographies of the confining walls are limited by our lithography methods, with rounding of sharper corners and a feature resolution of $\sim 2-3$ {\textmu}m (see Fig.~\ref{fig:experimentfigure1}(a), top row).

For the optical characterization, silica rods were imaged via transmission electron microscopy (Tecnai F20, Thermo Fisher Scientific) at $200$ kV, Fig.~\ref{fig:experiment2}(a). Samples were prepared by drop-casting ethanol-dispersed rods onto Formvar/Carbon 200 mesh copper grids ($\#$01801, Ted Pella). Fluorescently labeled rods were imaged using a Leica TCS SP8 STED 3X confocal microscope, Fig.~{1}(g-h), with a 63x/1.30 glycerol objective ($85$ wt\% glycerol in water). RITC-labeled rods were excited at $543$ nm using a pulsed super-continuum laser (SuperK, NKT Photonics), and emission was collected by a gated HyD detector ($0.3$–$6.0$ ns) over $554$–$691$ nm.

To examine the effect of topographical anchoring on defects in experiments, silica rods, Fig.~\ref{fig:experiment2}, are used to form a colloidal liquid crystal within circular confinement wells produced via contact photolithography. The confinement wells have topographies with a fixed periodicity of $p/D=10$ and five different amplitudes of $a/D=0.5$, 1, 2.5, 5, and 10, with $D$ the rod diameter. The confining walls of the circular wells are shown in Fig.~\ref{fig:experimentfigure1}(a). Note that the produced topographical features have lower amplitudes and smoothed features due to resolution limits of soft lithography. To form a colloidal liquid crystal, we concentrate the silica rods at the bottom of our samples via sedimentation, see experimental Video~7. The rod packing fraction in the solution is carefully controlled so that only a 1-3 rod layers form within each well. The measured rod packing fraction $\eta$ in experiments varies between 0.6 and 0.8 across all samples, with errors in $\eta$ of $\sim 0.1$ across samples due to imperfect particle tracking and variation in sedimentation times before measurement. The packing fraction is low enough to access both the nematic and smectic phase for our rods with aspect ratio $L/D\approx 9$~\cite{bates2000phase}. Nematic order and rod layering from smectic order are evident in experimental snapshots, shown in Fig.~\ref{fig:experimentfigure1}(b). Experimental Videos~1 and 2 show that the rods remain diffusive within the sample plane and can even exhibit bistable anchoring switching (experimental Video~4), proving that the rods are not kinetically trapped.

\subsection{Simulation}

\subsubsection{Monte Carlo simulations}

We model colloidal hard rods sedimented on a glass substrate as 2D hard ``discorectangles'' composed of a rectangle of length $L$ and width $D$, capped at both ends by semicircles of diameter $D$. In our investigation, we focus on systems of particles with length-to-diameter ratios $L/D=10,16$ and 25. In order to evaluate the shortest distance between a pair of rods and determine whether they overlap, we implement the algorithm by Vega and Lago~\cite{vega1994fast}. To capture the effects of confinement, either in rectangular or circular geometries, as detailed in the main text, we model the boundaries as impenetrable hard walls. Each rod $i$, located at position $\mathbf{r}_{i}$ and oriented along the unit vector $\hat{\mathbf{u}}_i=(\cos\phi_i,\sin\phi_i)^T$ (with $\phi_i$ the angle of the particle relative to a fixed axis), is represented by $L+1$ virtual sites distributed along $\mathbf{r}_{i} + \gamma \hat{\mathbf{u}}_i $, with $ |\gamma|<L/2 $. Confinement is enforced by applying the following external (boundary) potential to these sites:
\begin{equation}
    \beta\Phi_{w}(x)=
    \begin{cases}
 \displaystyle \infty & \text{for } x\leq 0, \\
\displaystyle 0 & \text{otherwise },
\end{cases}
\end{equation}
where $x$ denotes the minimal perpendicular distance from the virtual sites to the wall, with $x>0$ corresponding to the interior of the cavity. This potential effectively confines particles within the cavity boundaries, which can be either smooth, to promote planar anchoring of the rods, or patterned with triangular waves to induce homeotropic anchoring conditions. In the case of rectangular cavities with base $B$ and height $A$, we model the rough boundary along the top (short) side of the rectangle using a simple sinusoidal-triangular wave
\begin{equation}
\label{eq:wave}
    y(x)=y_0+\frac{2a}{\pi} \arcsin\left[ \sin\left(\frac{2\pi x}{p} \right) \right],
\end{equation}
where $y_0=A$, and $a$ and $p$ represent the amplitude and period (in units of length $D$) of the triangular wave. For circular cavities, we wrap the triangular wave around the perimeter of the circle. To achieve this, the effective radius of confinement as a function of the polar angle $\theta \in [0,2\pi)$ is given by
\begin{equation}
\label{eq:waveR}
    R(\theta) = R_0 + \frac{2a}{\pi}\arcsin\left[ \sin\left(\frac{2\pi R_0}{p} \theta \right) \right]
\end{equation}
where $R_0$ denotes the base (unmodulated) circle radius. For hybrid boundaries with alternating modulated and unmodulated arc segments, the full angular domain is partitioned into $N_a$ angular sectors or wedges, each corresponding to a distinct boundary condition.

We perform standard Metropolis~\cite{metropolis1953equation} Monte Carlo (MC) simulations at constant number of particles $N$ and temperature $T$ while controlling the area fraction $\eta$ of the confined particles. The area fraction is defined as $\eta=N\mathcal{A}_p/A_c$, where $\mathcal{A}_p=LD+\pi D^2/4$ is the area of a single hard discorectangle, and $A_c$ is the area of the confining geometry. Since our focus is on nematic ordering, we select packing fractions $\eta$ corresponding to stable nematic phases reported in previous studies of bulk systems of hard discorectangles with various length-to-diameter ratios $L/D$~\cite{bates2000phase}. Specifically, we use $\eta=0.55$ ($L/D=10$), $\eta=0.40$ ($L/D=16$), and $\eta=0.33$ ($L/D=25$). To reach the target packing fraction $\eta$, we start from a low-density configuration ($\eta\ll1$) placed within a scaled-up (augmented) version of the confining geometry, which is then gradually compressed. During this compression phase, only smooth (unmodulated) boundaries are employed. Once the target area fraction is reached, triangular-wave modulations are introduced on the rough boundaries by linearly increasing the amplitude $a$ (see Eqs.~\ref{eq:wave} and \ref{eq:waveR}) during a MC simulation, thereby gradually transforming the confinement from an unmodulated to a modulated geometry. More specifically, departing from $a=0$, small increments $\Delta a=a/4000$, with 8000 MC cycles at each ``instantaneous'' $a$ (total $3.2\times10^{7}$ MC cycles) are implemented. Each cycle consists of $N$ trial moves in which a randomly selected particle is either translated or rotated with equal probability. After the desired area fraction and boundary modulation are established, the system is equilibrated over at least $5\times10^7$ MC cycles. Prior to equilibration, the maximum displacement parameters, used to control the step sizes for both translational and rotational moves, are adjusted to maintain an acceptance rate close to the standard value of 0.3.

To characterize defect structures and reveal the orientational patterns of the confined liquid crystal, we use a spatially resolved, second-rank, symmetric and traceless nematic order tensor defined in 2D as
\begin{equation}
    Q_{2D}^k=\frac{1}{\sum_i \ell_i^k}\left< \sum_i \ell_{i}^k \left[2\left(\hat{\boldsymbol{u}}_i \otimes \hat{\boldsymbol{u}}_j \right) - \mathbb{I}_2 \right] \right>,
\end{equation}
where the index $k$ labels a specific subarea, $\ell_{i}^{k}$ is the length of particle $i$ lying within subarea $k$, $\mathbb{I}_2$ is the 2D identity tensor, and $\left< \cdot \right>$ denotes an ensemble average. The length weighting ensures that the contribution of each particle is properly scaled by its relative length within the subarea~\cite{Grlea2016}. A frame-independent measure of local alignment is provided by the scalar order parameter $S^k$, defined as the positive eigenvalue $\lambda_+^k$ of the nematic order tensor (with the other eigenvalue $\lambda_-^k=-\lambda_+^k$). This parameter ranges from $0$ for a fully disordered system to $1$ for perfect alignment. In our implementation, we find that using squared subareas of size $4D \times4D$ offers a good compromise between resolution and computational cost. The scalar order parameter is averaged over 2000 configurations, each sampled every 500 MC cycles. This relatively short sampling interval between configurations enables the localization of topological defects as regions of low $S^k$, which would otherwise be averaged out due to thermal fluctuations. Fluctuations in the positions of the topological defects can be observed in the sparse (relatively) low-$S$ regions shown in the figures throughout the manuscript. To facilitate visual identification of topological defects, we manually overlay open circles on selected instantaneous simulation configurations at the locations where such topological defects appear.

We note that in sampling the different anchoring configurations reported in Fig. 2, two independent runs yield indistinguishable results and reveal a robust bistable region (triangle markers) that does not exhibit hysteresis when the amplitude $a$ is increased (forward transformation) or decreased (reverse transformation) near the transition line, provided the changes are sufficiently small and gradual. In our case, we have verified that, by fixing the period at $p/D=4$ and cyclically varying $a/D$ between 1 (planar) and $4$ (homeotropic), using very slow linear transformations with increments $\pm\Delta(a/D)=(4-1)/4000$ and 8000 MC cycles at each $a$, the anchoring transition always occurs regardless of the direction of the geometrical transformation (as shown in simulation Video~8).

\subsubsection{Landau-de Gennes theory}

We employ the Landau-de Gennes (LdG) theory for colloidal hard rods as presented in Ref.~\cite{everts2016landau}. We determine the nematic order parameter tensor $\mathbf{Q}(x,y)$ for a system of hard rods confined to a plane and varying confinements by minimizing the LdG grand potential using the Euler-Lagrange equations
\begin{eqnarray}
    (l_1 + l_s) \mathbf{\nabla}^2 \mathbf{q}& = & \nonumber \\
    & & \hspace{-18mm} 3 a \beta(\mu^* - \mu) \mathbf{q} - \frac{9}{2}b \mathbf{q} + \frac{d}{2} (3 + 16q_1^2+16q_2^2)\mathbf{q},
\end{eqnarray}
with the boundary condition
\begin{equation}
    (l_1+l_s)(\hat{\mathbf{v}}\cdot \mathbf{\nabla} \mathbf{q}) + \frac{9}{4} w (\mathbf{q} - \mathbf{q}^0) = 0 \\,
\end{equation}
where we use the dimensionless Landau coefficients $a=1.436$, $b=5.851$, and $d=3.693$, the chemical potential $\beta \mu=2.7$, the chemical potential at the isotropic spinodal $\beta\mu^*=6.855$, the elastic constants $l_1=0.165L^2$ and $l_s=0.837L^2$, and the anchoring strength $w=10L$, where we follow Ref.~\cite{everts2016landau}. The chemical potential considered results in low values of the nematic order, but it facilitates the occurrence of nematic defects across all geometries. Furthermore, $\mathbf{v}$ denotes the outward-pointing normal vector, and $\mathbf{q}$ determines the symmetric, traceless, 2D nematic order parameter tensor
\begin{equation}
    \mathbf{Q}(x,y) =
    \begin{pmatrix}
        \frac{1}{4} + q_1(x,y) & q_2(x,y) & 0 \\
        q_2(x,y) & \frac{1}{4} - q_1(x,y) & 0 \\
        0 & 0 & -\frac{1}{2}
    \end{pmatrix}
    \\.
\end{equation}
The tensor $\mathbf{q}^0$ determines the nematic order parameter tensor at the boundary, that is parallel (perpendicular) to the boundary for planar (homeotropic) anchoring. These equations were minimized using COMSOL Multiphysics\textregistered~software~\cite{comsol}.

\bibliography{library}

\end{document}